\documentclass[aps,prl,preprint,showpacs,groupedaddress]{revtex4}
\usepackage {amssymb}
\usepackage {amsmath}
\usepackage {graphicx}
\usepackage {longtable}

\begin{document}
\title{Antiferromagnetic fluctuations in the superconducting phase
of low- and high-temperature superconductors}

\author{Fedor V.Prigara}
\affiliation{Institute of Microelectronics and Informatics,
Russian Academy of Sciences,\\ 21 Universitetskaya, Yaroslavl
150007, Russia} \email{fvprigara@rambler.ru}

\date{\today}

\begin{abstract}

Based on recent experimental results for electron-doped cuprate oxides and
ferromagnetic superconductors, it is shown that antiferromagnetic
fluctuations always develop in the superconducting phase of both low- and
high-temperature superconductors. The relation between the magnitude of the
antiferromagnetic pseudogap and the characteristic temperature of the
antiferromagnetic pseudogap opening is obtained. The characteristic
temperature of the antiferromagnetic pseudogap opening for metal
superconductors is estimated.

\end{abstract}

\pacs{74.25.Ha, 74.25.Dw, 74.62.Fj, 64.60.-i}

\maketitle

Recently, it was shown by means of inelastic neutron diffraction
measurements [1] that earlier discovered in electron-doped cuprate
oxides $R_{2 - x} Ce_{x} CuO_{4} $, where R= Sm, Nd, Pr, (in the
angle resolved photoemission measurements) [2] the
antiferromagnetic pseudogap $\Delta ^{\ast }$ corresponds to
antiferromagnetic fluctuations in the superconducting phase of
this cuprate superconductors (for a review see Ref. 3). The
magnitude $\Delta ^{ *} $ of the antiferromagnetic pseudogap is
much larger than the magnitude $\Delta $ of the superconducting
gap, and the characteristic temperature $T_{AFM}^{ *}  $ of the
antiferromagnetic pseudogap opening essentially exceeds the
critical temperature $T_{c} $ of the superconducting transition.
Such a behavior is typical for high-temperature superconductors,
if normally observed in them above $T_{c} $ pseudogap $\Delta ^{
*} $ can be identified with the antiferromagnetic pseudogap
corresponding to antiferromagnetic fluctuations with a finite
lengthscale. Here we show that antiferromagnetic fluctuations
develop in the superconducting phase of both low- and
high-temperature superconductors and estimate the magnitude
$\Delta ^{ *} $ of the antiferromagnetic pseudogap, the
characteristic length scale \textit{l} of antiferromagnetic
fluctuations, and also, for metal superconductors, the
characteristic temperature $T_{AFM}^{ *}  $ of the
antiferromagnetic pseudogap opening.

Recent detailed studies of the AC magnetic susceptibility $\chi
\left( {T} \right)$ as a function of the temperature \textit{T} in
the ferromagnetic superconductor $ErRh_{4} B_{4} $ with the
critical temperature $T_{c} = 8.7K$ and the Curie temperature
$T_{FM} = 0.97K$ [4] have revealed a maximum of $\chi \left( {T}
\right)$ at the temperature $T_{AFM}^{ *}  \cong 2K$ (in fields $H
\cong 3kOe$), which can be interpreted as the manifestation of
strong antiferromagnetic fluctuations in the superconducting phase
of this low-temperature superconductor. This interpretation is
supported by the sequence of phase transitions in the other
ferromagnetic superconductor $ErNi_{2} B_{2} C$: the ferromagnetic
transition with the transition temperature $T_{FM} = 2.3K$, the
antiferromagnetic transition with the transition temperature
$T_{AFM} = 6K$, and the superconducting transition with $T_{c} =
11K$.

Thus, we have similar sequences of transitions:

\newcommand{\PreserveBackslash}[1]{\let\temp=\\#1\let\\=\temp}
\let\PBS=\PreserveBackslash
\begin{longtable}
{|p{106pt}|p{106pt}|p{106pt}|p{106pt}|}
a & a & a & a  \kill
\hline
\[
ErRh_{4} B_{4}
\]
&
\[
T_{FM} = 0.97K
\]
&
\[
T_{AFM}^{ *}  \cong 2K
\]
&
\[
T_{c} = 8.7K
\]
 \\
\hline
\textbf{}$ErNi_{2} B_{2} C$&
\[
T_{FM} = 2.3K
\]
&
\[
T_{AFM} = 6K
\]
&
\[
T_{c} = 11K
\]
 \\
\hline
\end{longtable}

\textbf{}

These results suggest that antiferromagnetic fluctuations are
present in the superconducting phase of conventional,
low-temperature superconductors. (The above ferromagnetic
superconductor $ErRh_{4} B_{4} $ is type I superconductor, as it
is shown in Ref.4). However, the characteristic temperature
$T_{AFM}^{ *}  $ of the antiferromagnetic pseudogap opening is, in
the case of low-temperature superconductors, less than the
critical temperature $T_{c} $ of the superconducting transition
(contrary to the case of high-temperature cuprate
superconductors).

Recently, it was shown [5] that the melting of a crystalline solid
occurs when the critical number density $n_{c} $ of vacancies is
achieved, the critical number density of vacancies being

\begin{equation}
\label{eq1}
n_{c} \cong n_{0} exp\left( { - \alpha}  \right),
\end{equation}

\noindent where $n_{0} \approx 1.1 \times 10^{22}cm^{ -
3}$\textbf{} and $\alpha \approx 18$\textbf{} are quantum
constants [6]. The length scale

\begin{equation}
\label{eq2}
d_{c} = n_{c}^{ - 1/3} \cong a_{0} exp\left( {\alpha /3} \right) \cong
180nm,
\end{equation}

\noindent where $a_{0} = n_{0}^{ - 1/3} \cong 0.45nm$, can be
interpreted as the size of a crystalline domain which cannot
contain more than one vacancy. This interpretation is supported by
the experimental results on the recryctallization of thin Al foils
[7]. The recrystallization occurs only if the foil thickness
\textit{d} exceeds the length scale $d_{c} $, in more thin films
there were no additional vacancies to nucleate the
recrystallization [7].

This concept can be extended on other phase transitions, such as
ferroelectric, ferromagnetic, and superconducting, if we consider the
corresponding phase transition as the melting of some lattice. In the case
of the ferroelectric transition, it is the lattice of electrical dipoles; in
the case of the ferromagnetic transition, it is the lattice of magnetic
dipoles. In the case of the superconducting phase transition, the situation
is more complex, since the electron degrees of freedom are essentially
involved. However, as it will be shown below, the above concept holds for
superconducting transitions too.

The thermodynamic consideration based on the Clausius- Clapeyron
equation gives the number density $n_{v} $\textit{} of vacancies
in a solid in the form [6]

\begin{equation}
\label{eq3}
n_{v} = \left( {P_{0} /T} \right)exp\left( { - E_{v} /T} \right) = \left(
{n_{0} T_{0} /T} \right)exp\left( { - E_{v} /T} \right),
\end{equation}

\noindent
where $E_{v} $ is the energy of the vacancy formation, $P_{0} = n_{0} T_{0}
$ is a constant, $T_{0} $ can be put equal to the melting temperature of the
solid at ambient pressure, and the Boltzmann constant $k_{B} $ is included
in the definition of the temperature \textit{T}.

The energy of the vacancy formation $E_{v} $ depends linearly on the
pressure \textit{P} (in the region of high pressures) as given by the
formula

\begin{equation}
\label{eq4}
E_{v} = E_{v0} - \alpha P/n_{0} ,
\end{equation}

\noindent
where $\alpha $ is a dimensionless constant, $\alpha \approx 18$ for
sufficiently high pressures. On the atomic scale, the pressure dependence of
the energy of the vacancy formation in the equation (\ref{eq4}) is produced by the
strong atomic relaxation in a crystalline solid under high pressure.

According to the equations (\ref{eq3}) and (\ref{eq4}), the number density of vacancies in a
solid increases with increasing pressure (in the region of high pressures),

\begin{equation}
\label{eq5}
n_{v} = \left( {n_{0} T_{0} /T} \right)exp\left( { - \left( {E_{v0} - \alpha
P/n_{0}}  \right)/T} \right).
\end{equation}

Consider now the number density \textit{n} of elementary excitations in a
corresponding lattice, instead of the number density of vacancies in an
ordinary lattice. An elementary excitation invokes the change of the sign of
the electrical or magnetic dipole moment of the elementary lattice cell for
ferroelectrics and ferromagnets, respectively. In the case of
superconductors, an elementary excitation is more complex.

Similarly to the energy of the vacancy formation (\ref{eq4}), the energy \textit{E}
of an elementary excitation depends on the pressure (in the region of high
pressures) as follows

\begin{equation}
\label{eq6}
E = E_{0} - \alpha _{P} P/n_{0} ,
\end{equation}

\noindent
where $E_{0} $ is the energy of an elementary excitation at ambient
pressure, and $\alpha _{P} $ is a dimensionless constant dependent on the
type of the phase transition. The pressure dependence of the energy of an
elementary excitation in the equation (\ref{eq6}) is produced by the atomic
relaxation in the corresponding lattice under high pressure.

The pressure dependence of the number density \textit{n} of elementary
excitations is given by the formula analogous to the equation (\ref{eq5}),

\begin{equation}
\label{eq7}
n = \left( {n_{0} T_{c} /T} \right)exp\left( { - \left( {E_{0} - \alpha _{P}
P/n_{0}}  \right)/T} \right),
\end{equation}

\noindent
where $T_{c} $ denotes here the transition temperature at ambient pressure.

The corresponding phase transition occurs when the critical number
density $n_{c} $ of elementary excitations (one elementary
excitation per crystalline domain) is achieved. (Inelastic neutron
diffraction measurements [1] have revealed that, on cooling, the
antiferromagnetic phase transition occurs when the characteristic
length \textit{l} of antiferromagnetic fluctuations achieves the
size of a crystalline domain given by the equation (\ref{eq2}), $l
\cong 400a_{0} $.) In view of the equation (\ref{eq7}), it means
that, on the phase equilibrium curve in the high pressure region,
we have

\begin{equation}
\label{eq8}
\left( {E_{0} - \alpha _{P} P/n_{0}}  \right)/T \cong E_{0} /T_{c} \cong
\alpha .
\end{equation}

The equation (\ref{eq8}) gives the pressure dependence of the transition temperature
\textit{T} in the region of high pressures in the form

\begin{equation}
\label{eq9}
 T \cong T_{c} - \left( {\alpha _{P} /\alpha}  \right)P/n_{0} .
\end{equation}

The dimensionless slope of the phase equilibrium curve is given by
the formula

\begin{equation}
\label{eq10} k = n_{0} dT/dP \cong - \alpha _{P} /\alpha .
\end{equation}

Similarly to the case of the melting of crystalline solids, the
relations (\ref{eq9}) and (\ref{eq10}) are valid only for
sufficiently high pressures. In the region of relatively low
pressures, the sign of $dT/dP$ can be positive (the transition
temperature may increase with pressure).

The superconducting phase diagram of Li metal [8] gives $k \approx
- 1/96$ and $\alpha _{P} \approx - k\alpha \approx 3/16$ for
sufficiently high pressures, if we assume the relation ${P}'
\approx 200P$ between the real pressure \textit{P} and the
measured pressure ${P}'$. (The problem of the calibration of high
pressures is considered in Ref.6). These values of \textit{k} and
$\alpha _{P} $ are characteristic for low-temperature
superconductors (see data for Al, Sn, Pb in Ref.9).

The superconducting phase diagram of the cuprate oxide $La_{1.48}
Nd_{0.4} Sr_{0.12} CuO_{4} $ [10] gives $k \approx - 1/48$ and
$\alpha _{P} \approx - k\alpha \approx 3/8$, for sufficiently high
pressures. These values of \textit{k} and $\alpha _{P} $ are
characteristic for high-temperature superconductors (see, for
example, data for $MgB_{2} $ in Ref. 11 and data for $Rb_{3}
C_{60} $ in Ref. 9).

The values of \textit{k} and $\alpha _{P} $ for ferroelectrics are
$k \approx - 1/9$ and $\alpha _{P} \approx - k\alpha \approx 2$
(see data for $BaTiO_{3} $ in Ref. 12). In the case of
ferromagnets, $k \approx - 2/9$ and $\alpha _{P} \approx - k\alpha
\approx 4$ (see data for the weak itinerant ferromagnet $SrRuO_{3}
$ in Ref. 13).

In accordance with the equation (\ref{eq8}), the energy $E_{0} $ of an elementary
excitation at ambient pressure is given by the relation

\begin{equation}
\label{eq11} E_{0} \cong \alpha T_{c} .
\end{equation}

In the case of $MgB_{2} $ with $T_{c} = 39K$, the energy of an
elementary excitation has an order of the Debye temperature
$\theta _{D} $: $E_{0} \cong \alpha T_{c} \cong 700K$, whereas
$\theta _{D} = 800 \pm 80K$ [14, 15, 16].

If we assume that at $T \cong 0$ the transition pressure $P_{c} $
has an order of the melting pressure $P_{m} \cong n_{0} T_{m} $,
where $T_{m} $ is the melting temperature of the solid at ambient
pressure [6], then

\begin{equation}
\label{eq12} \alpha _{P} \cong n_{0} E_{0} /P_{c} \cong E_{0}
/T_{m} \cong \theta _{D} /T_{m} ,
\end{equation}

\noindent
for superconductors with $E_{0} \cong \theta _{D} $. For most of metals, the
ratio of the Debye temperature to the melting temperature is $\theta _{D}
/T_{m} \cong 0.2 - 0.4$.

The magnitude $2\Delta ^{ *} $ of the antiferromagnetic pseudogap
is equal to the energy $E_{0} $ of an elementary antiferromagnetic
excitation, so that, in accordance with the equation (\ref{eq11}),
we have

\begin{equation}
\label{eq13} 2\Delta ^{ *}  \cong \alpha T_{AFM}^{ *}  \quad ,
\end{equation}

\noindent where $T_{AFM}^{ *}  $ is the characteristic temperature
of the antiferromagnetic pseudogap opening. Optical measurements
in $Nd_{2 - x} Ce_{x} CuO_{4} $ and $Pr_{2 - x} Ce_{x} CuO_{4} $
give the relation $2\Delta ^{ *} /T_{AFM}^{ *}  \approx 16$ [3],
which is close to the relation (\ref{eq12}), since $\alpha \approx
18$.

The experimental data suggest that the magnitude $2\Delta $ of the
superconducting gap is given by the relation

\begin{equation}
\label{eq14} 2\Delta \cong \alpha _{P} \alpha T_{c} .
\end{equation}

For low-temperature superconductors $\alpha _{P} \approx - k\alpha
\approx 3/16$ and the equation (\ref{eq14}) gives

\begin{equation}
\label{eq15} 2\Delta \cong 3.4T_{c} .
\end{equation}

This estimation of the magnitude of the superconducting gap is
consistent with experimental data for Al, In, Sn obtained in the
tunneling experiments [17]. Other experimental techniques give
slightly different ratios of the magnitude of the superconducting
gap to the critical temperature, maybe due to the interplay
between the superconducting gap and the antiferromagnetic
pseudogap (see below).

For high-temperature superconductors $\alpha _{P} \approx -
k\alpha \approx 3/8$ and from the equation (\ref{eq14}) we obtain

\begin{equation}
\label{eq16} 2\Delta \cong 7T_{c} .
\end{equation}

The ratio $2\Delta /T_{c} $ measured in $Sm_{2 - x} Ce_{x} CuO_{4}
$ by means of local tunneling spectroscopy is varying from 4 to 7
due to the local change in cerium content and, whence, in the
value of the critical temperature [1]. The maximum value of
$2\Delta /T_{c} $, corresponding to the maximum value of the
critical temperature (which was the case in optimally doped
samples used in these measurements) is in agreement with the
relation (\ref{eq16}).

The characteristic length \textit{l} of antiferromagnetic
fluctuations in the superconducting phase [3] has an order of
magnitude of the diameter $2r_{0} $ of the atomic relaxation
region [5],

\begin{equation}
\label{eq17} l \cong 2r_{0} \cong 2\alpha a_{0} \cong 16nm,
\end{equation}

\noindent
or, more exactly,

\begin{equation}
\label{eq18} l \cong 4\beta \alpha r_{0} \cong 25nm,
\end{equation}

\noindent
if $\beta \cong 0.8$ wich is the case for most of metals [3].

Inelastic neutron diffraction measurements give the characteristic
length of antiferromagnetic fluctuations in the superconducting
phase of $R_{2 - x} Ce_{x} CuO_{4} $ at the level of $l \approx
20nm$ [1,3], which is in agreement with the estimations
(\ref{eq17}) and (\ref{eq18}). It means that antiferromagnetic
fluctuations are connected with the atomic displacement, the
region of antiferromagnetic fluctuations having its center at the
position of the ``lattice defect'' associated with an elementary
antiferromagnetic excitation.

Neutron diffraction studies of the ferromagnetic superconductor
$ErRh_{4} B_{4} $ have established the existence of a modulated
structure in the superconducting regions at the length scale $
\cong 10nm$ [4]. This structure can be interpreted as the
antiferromagnetic modulation with a length scale having an order
of the radius of the atomic relaxation region $r_{0} \cong \alpha
a_{0} \cong 8nm$ [5]. The size of superconducting regions of the
supercooled superconducting phase of this ferromagnetic
superconductor, coexisting with ferromagnetic domains in the
narrow temperature interval $T \approx 0.9 - 0.97K$, has an order
of the size of crystalline domains given by the equation
(\ref{eq2}). On heating, a sharp transition from the ferromagnetic
phase to the superconducting phase is observed at $T_{FM} = 0.97K$
[4]. The ratio $T_{c} /T_{FM} $ for this ferromagnetic
superconductor is $T_{c} /T_{FM} \cong \alpha /2 \cong 9$.

Since $T_{AFM}^{ *}  \cong 2K$ and $T_{c} = 8.7K$, from the
relation (\ref{eq13}) we obtain for $ErRh_{4} B_{4} $:

\begin{equation}
\label{eq19}
2\Delta ^{ *}  \cong \alpha T_{AFM}^{ *}  \cong 36K,
\end{equation}

\noindent whereas the magnitude of the superconducting gap,
according to the relation (\ref{eq15}), is

\begin{equation}
\label{eq20}
2\Delta \cong 3.4T_{c} \cong 30K,
\end{equation}

\noindent so that $2\Delta ^{ *}  \cong 2\Delta $. For the other
ferromagnetic superconductor $ErNi_{2} B_{2} C$, the equations
(\ref{eq13}) and (\ref{eq15}) give $2\Delta ^{ *}  \cong 110K$ and
$2\Delta \cong 37K$, respectively, so that $\Delta ^{ *} /\Delta
\cong 3$.

Since $2\Delta ^{ *}  \cong 2\Delta $ for $ErRh_{4} B_{4} $, the
antiferromagnetic phase transition is suppressed, contrary to the
case of $ErNi_{2} B_{2} C$, where $\Delta ^{ *} /\Delta \cong 3$.
In the case of weak itinerant ferromagnets $UGe_{2} $ and
\textit{URhGe} [2], the relation $2\Delta ^{ *}  \cong 2\Delta $
is also presumably valid, so that the antiferromagnetic transition
is suppressed and the superconducting transition occurs instead.
The characteristic temperature of the antiferromagnetic pseudogap
opening can be determined from the relations (\ref{eq13}) and
(\ref{eq14}) as follows

\begin{equation}
\label{eq21} T_{AFM}^{ *}  \cong 2\Delta ^{ *} /\alpha \cong
2\Delta /\alpha \cong \alpha _{P} T_{c} .
\end{equation}

Thus we obtain for these superconducting ferromagnets

\begin{longtable}
{|p{106pt}|p{106pt}|p{106pt}|p{106pt}|}
a & a & a & a  \kill
\hline
\[
UGe_{2}
\]
&
\[
T_{AFM}^{ *}  \cong 0.2K
\]
&
\[
T_{c} = 0.95K
\]
 \par (under pressure)&
\[
T_{FM} = 33K
\]
 \\
\hline
\textbf{}\textit{URhGe}&
\[
T_{AFM}^{ *}  \cong 0.05K
\]
&
\[
T_{c} = 0.27K
\]
&
\[
T_{FM} = 9.5K
\]
 \\
\hline
\end{longtable}

\textbf{}

The ratio $T_{FM} /T_{c} $ for these ferromagnetic superconductors is
$T_{FM} /T_{c} \cong 2\alpha $.

Since the antiferromagnetic transition is not normally observed in metal
superconductors, the same relations $2\Delta ^{ *}  \cong 2\Delta $ and
$T_{AFM}^{ *}  \cong \alpha _{P} T_{c} $ are valid in this case too.

The $^{17}$O nuclear magnetic resonance spectra of overdoped
samples of the hole-doped high-temperature superconductor
$Bi_{2}Sr_{2}CaCu_{2}O_{8+\delta}$ with $T_{c}=75-90K$ [18] show a
minimum of the linewidth of the O(1) resonance (corresponding to
the oxygen in the $CuO_{2}$ plane) at the temperature $T_{AFM}^{
*} \cong \alpha _{P} T_{c} $, where $\alpha _{P} \approx 3/8$.
This minimum of the O(1) resonance linewidth corresponds to
antiferromagnetic fluctuations in the superconducting phase of
this cuprate superconductor and indicates that the relation
$2\Delta ^{ *} \cong 2\Delta $ is valid also for overdoped cuprate
superconductors. In the case of near-optimally doped sample with
$T_{c}=90K$ the increase of the Knight shift above $T_{c}$ is
indicative of the relation $2\Delta ^{ *}  > 2\Delta $ in the
region of optimal doping.

To summerize, we show that antiferromagnetic fluctuations develop
in the superconducting phase of both low- and high-temperature
superconductors, we obtain the temperature and pressure dependence
of the number density of elementary excitations for
ferroelectrics, ferromagnets and superconductors. We obtain
further the curve of phase equilibrium in the high pressure region
for ferroelectric, ferromagnetic and superconducting phase
transitions. We give also the estimations of the magnitudes of the
antiferromagnetic pseudogap and the superconducting gap, and of
the characteristic length scale of antiferromagnetic fluctuations
in the superconducting phase. We show that for metal
superconductors (and other nonmagnetic low-temperature
superconductors) the magnitudes of the antiferromagnetic pseudogap
and the superconducting gap are close each to other and give the
estimation of the characteristic temperature of the
antiferromagnetic pseudogap opening for these superconductors.

\begin{center}
---------------------------------------------------------------
\end{center}

[1] E.M.Motoyama, G.Yu, I.M.Vishik, O.P.Vajk, P.K.Mang, and
M.Greven, Nature \textbf{445}, 186 (2007).

[2] N.P.Armitage, D.H.Lu, C.Kim et al., Phys. Rev. Lett.
\textbf{87}, 147003 (2001).

[3] A.Zimmers, Y.Noat, T.Cren, W.Sacks, D.Roditchev, B.Liang, and
R.L.Greene, E-print archives, arXiv: 0705.4217 (cond-mat.
supr-con) (2007).

[4] R.Prozorov, M.D.Vannette, S.A.Law, S.L.Bud'ko, and
P.C.Canfield, E-print archives, arXiv: 0707.0029 (cond-mat.
str-el) (2007).

[5] F.V.Prigara, E-print archives, arXiv: 0704.0972 (cond-mat.
stat-mech) (2007).

[6] F.V.Prigara, E-print archives, cond-mat/0701148 (2007).

[7] B.S.Bokstein, S.Z.Bokstein, and A.A.Zhukhovitsky,
\textit{Thermodynamics and Kinetics of Diffusion in Solids}
(Metallurgiya Publishers, Moscow, 1974).

[8] S.Deemyad and J.S.Schilling, Phys. Rev. Lett. \textbf{91},
167001 (2003).

[9] J.S.Schilling, in \textit{Handbook of High Temperature
Superconductivity: Theory and Experiment}, editor J.R.Schrieffer,
associated editor J.S.Brooks (Springer Verlag, Hamburg, 2007).

[10] M.K.Crawford, R.L.Harlow, S.Deemyad et al., Phys. Rev. B
\textbf{71}, 104513 (2005).

[11] S.Deemyad, T.Tomita, J.J.Hamlin et al., Physica C
\textbf{385}, 105 (2003).

[12] G.S.Zhdanov, \textit{Solid State Physics} (Moscow University
Press, Moscow, 1961).

[13] J.J.Hamlin, S.Deemyad, J.S.Schilling et al., E-print
archives, cond-mat/0609149 (2006).

[14] Z.F.Wei, G.C.Che, F.M.Wang, M.He, and X.L.Chen, Modern
Physics Lett. B \textbf{15}, 1109 (2001).

[15] T.Ichitsubo, H.Ogi, S.Nishimura, T.Seto, M.Hirao, and H.Inui,
Phys. Rev. B \textbf{66}, 052514 (2002).

[16] J.L.Luo, J.Zhang, ZJ.Chen et al., Chinese Physics Lett.
\textbf{18}, 820 (2001).

[17] D.Saint-James, G.Sarma, and E.J.Thomas, \textit{Type II
Superconductivity} (Pergamon Press, Oxford, 1969).

[18] B.Chen, S.Mukhopadhyay, W.P.Halperin, P.Guptasarma, and
D.G.Hinks, E-print archives, arXiv:0707.3171 (cond-mat.supr-con)
(2007).

\end{document}